\documentclass[12pt,preprint]{aastex}

\shorttitle{INTERSTELLAR LINES TOWARD NGC 5195}
\shortauthors{RITCHEY \& WALLERSTEIN}

\begin{document}
\title{The Interstellar Line of Sight to the Interacting Galaxy NGC 5195\footnote{Based on observations obtained with the Apache Point Observatory 3.5-m telescope, which is owned and operated by the Astrophysical Research Consortium.}}
\author{Adam M. Ritchey\altaffilmark{1} and George Wallerstein\altaffilmark{1}}
\altaffiltext{1}{Department of Astronomy, University of Washington, Seattle, WA 98195, USA; aritchey@astro.washington.edu}

\begin{abstract}
We present moderately-high resolution echelle observations of the nucleus of NGC~5195, the line of sight to which samples intervening interstellar material associated with the outer spiral arm of M51. Our spectra reveal absorption from interstellar Na~{\sc i}, K~{\sc i}, Ca~{\sc ii}, and CH$^+$, and from a number of diffuse interstellar bands (DIBs), at a velocity close to that exhibited by H~{\sc i}~21~cm emission from M51 at the position of NGC~5195. The H~{\sc i} column density implied by the equivalent width of the $\lambda5780.5$ DIB, based on the relationship between $W_{\lambda}$(5780.5) and $N$(H~{\sc i}) derived for sight lines in the local Galactic interstellar medium, is consistent with the column density obtained from the integrated H~{\sc i} emission. The H$_2$ column density predicted from the observed column density of K~{\sc i}, using the Galactic relationship between $N$(K~{\sc i}) and $N$(H$_2$), is comparable to $N$(H~{\sc i}), suggesting a high molecular fraction ($\sim$0.65) for the M51 gas toward NGC~5195. The DIBs toward NGC~5195 are, on average, $\sim$40\% weaker than would be expected based on the K~{\sc i} column density, a further indication that the gas in this direction has a high molecular content. The M51 material is characterized also by a high $N$(Na~{\sc i})/$N$(Ca~{\sc ii}) ratio ($\gtrsim11$), indicative of a high degree of Ca depletion, and a high $W_{\lambda}$(5797.1)/$W_{\lambda}$(5780.5) ratio ($\sim$1.6), suggestive of either a very weak ambient radiation field or a significantly shielded environment. A high $N$(CH$^+$)/$N$(CH) ratio ($\gtrsim2.3$) for the M51 material toward NGC~5195 may be the result of enhanced turbulence due to interactions between M51 and its companion.
\end{abstract}


\section{INTRODUCTION}
The grand design spiral galaxy M51 (NGC~5194) has long been known to be interacting with its companion, the SB0 galaxy NGC~5195, which is seen in projection behind dusty interstellar gas now understood to be associated with the northeastern spiral arm of M51. Spinrad (1973) commented on the obvious dust lanes that are seen in front of NGC~5195, and speculated that some of these could be due to foreground absorption from the outer arm of M51. However, it was Van Dyk (1987) who confirmed this speculation, discovering anomalously strong Na~{\sc i}~D lines attributable to interstellar gas in the arm of M51 projected onto the nucleus of NGC~5195. The strength of the interstellar features combined with the brightness of the background source makes the line of sight to NGC~5195 ideal for studying the interstellar medium (ISM) of a nearby ($d\sim8.0$ Mpc; Karachentsev et al.~2004) interacting galaxy using the technique of absorption-line spectroscopy. The low resolution of the spectrum analyzed by Van Dyk (1987), however, precluded a detailed investigation of the ISM in this direction.

In Ritchey \& Wallerstein (2012), we presented moderately-high resolution echelle observations of SN~2011dh in M51, which revealed numerous interstellar absorption components in Na~{\sc i}~D and Ca~{\sc ii}~H and K associated with the supernova host galaxy. The M51 components were found to span a velocity range of over 140 km~s$^{-1}$, making them similar to intermediate-velocity clouds observed in the halo of the Milky Way. Yet many of the M51 components toward SN~2011dh, even those at extreme positive and negative velocities, were found to have Na~{\sc i}/Ca~{\sc ii} column density ratios of $\sim$1.0. Such values are typically associated with relatively cool quiescent gas in the disk of a galaxy like M51 or the Milky Way. Some of these components may thus be tracing gas that has been tidally stripped from the disk of M51 or its companion (or both) during one of the multiple encounters which are thought to have occurred recently between the two galaxies (e.g., Salo \& Laurikainen 2000; Theis \& Spinneker 2003).

Background sources suitable for probing the ISM of nearby galaxies that are bright enough for high-resolution absorption-line spectroscopy are rare. Extragalactic studies using individual stars as background sources are currently limited to examining sight lines within Local Group galaxies, such as the Large and Small Magellanic Clouds (LMC and SMC; e.g., Cox et al.~2006, 2007; Welty et al.~2006; Welty \& Crowther 2010), M31 (e.g., Cordiner et al.~2008a, 2011), and M33 (e.g., Cordiner et al.~2008b; Welsh et al.~2009). Bright supernovae are perhaps the best sources for probing the ISM of galaxies beyond the Local Group, and, indeed, many nearby galaxies have been investigated through supernova absorption-line spectroscopy (e.g., Wallerstein et al.~1972; D'Odorico et al.~1989; Bowen et al.~1994, 2000; Vladilo et al.~1994; Ho \& Filippenko 1995; Sollerman et al.~2005; Cox \& Patat 2008, 2014; Welty et al.~2014; Ritchey et al.~2015). However, supernovae fade over time, making follow-up high-resolution observations difficult, and the occurrence of any single supernova cannot be predicted. The nuclei of nearby spiral galaxies can be relatively bright, but often do not show measureable interstellar absorption features (unless the galaxy is observed in an edge-on orientation). Thus, the chance position of NGC~5195 located behind a prominant spiral arm of M51 presents a unique situation.

In this paper, we examine moderately-high resolution echelle observations of the nucleus of NGC~5195, which allow us to study the properties of the interstellar gas seen in absorption along the line of sight. The spectra reveal redshifted interstellar absorption features of Na~{\sc i}, K~{\sc i}, Ca~{\sc ii}, and CH$^+$, along with a number of diffuse interstellar bands (DIBs), all of which help to constrain the physical conditions in the intervening interstellar material in front of NGC~5195. We briefly examine some of the prominent emission lines and stellar absorption lines in the spectrum of NGC~5195, and in spectra of the nucleus of M51, in order to place our measurements of the interstellar features in the proper context. In particular, a comparison of the radial velocities of the various absorption and emission features allows us to unambiguously associate the redshifted interstellar absorption lines toward NGC~5195 with gas in the spiral arm of M51.

\section{OBSERVATIONS AND DATA REDUCTION}
Our observations were acquired using the Astrophysical Research Consortium (ARC) echelle spectrograph on the 3.5 m telescope at Apache Point Observatory (APO). The echelle provides complete wavelength coverage in the range 3800--10200 \AA{} with a resolving power of $R\approx31,500$ ($\Delta v\approx9.5$~km~s$^{-1}$). A total of eighteen 30 minute exposures of the nucleus of NGC~5195 ($V=9.55\pm0.07$; Gil de Paz et al.~2007) were obtained between 2012 January and 2013 April. In the earliest spectra that we obtained, it was apparent that the Na~{\sc i}~D features at the redshift of M51 were quite strong in this direction, while the Ca~{\sc ii}~H and K lines were buried in the noise. Since the Na~{\sc i}/Ca~{\sc ii} column density ratio can be a useful diagnostic of physical conditions, we continued to accumulate observations to build up the signal until the Ca~{\sc ii} lines were clearly visible. On most occasions, the nearby, early-type star $\eta$~UMa (B3 V; $V=1.85$) was observed on the same night as NGC~5195 to aid in the removal of telluric absorption lines, which affect the Na~{\sc i}~D and K~{\sc i}~$\lambda\lambda7664,7698$ features. We obtained two 30 minute exposures of the nucleus of M51 ($V=8.36\pm0.06$; Gil de Paz et al.~2007) in 2012 January for comparison with NGC~5195.

Reduction of the raw data was accomplished using a semi-automated pipeline reduction script, which employs standard IRAF\footnote{IRAF is distributed by the National Optical Astronomy Observatory, which is operated by the Association of Universities for Research in Astronomy (AURA) under cooperative agreement with the National Science Foundation.} routines for bias correction, cosmic-ray removal, scattered-light subtraction, one-dimensional spectral extraction, flat-fielding, and wavelength calibration. The atmospheric absorption lines near the Na~{\sc i}~D and K~{\sc i} features were removed from the individual, calibrated exposures using the {\sc telluric} task within IRAF. We find that this procedure is effective even for removing the strong absorption lines associated with the $A$ band of atmospheric O$_2$ in the vicinity of the K~{\sc i} doublet, which normally hinder the analysis of the K~{\sc i}~$\lambda7664$ line. For these observations, we are aided also by the redshift of M51, which places the associated K~{\sc i}~$\lambda7664$ feature in a region mostly free from absorption by atmospheric O$_2$. After correcting for telluric absorption lines, all of the exposures were shifted to the reference frame of the local standard of rest (LSR) and co-added to produce a final, high signal-to-noise ratio (S/N) spectrum. Any overlapping portions of the echelle orders containing absorption (or emission) lines of interest were combined to improve the S/N and the resulting spectra were normalized to the continuum via low-order polynomial fits.

The S/N (per pixel) in the final, co-added spectrum of NGC~5195 peaks at $\sim$200 near the K~{\sc i} doublet, and decreases to $\sim$150 near the Na~{\sc i}~D features. At the Ca~{\sc ii}~H and K lines, the S/N is only $\sim$15, while near the Ca~{\sc ii} infrared triplet, the S/N is $\sim$115. For comparison, the S/N in our spectrum of the nucleus of M51, which had a total integration time of only 60 minutes, peaks at $\sim$60 near K~{\sc i} and decreases to $\sim$10 at Ca~{\sc ii}~K. Figure~1 presents a portion of the final, co-added spectrum of NGC~5195 in the vicinity of Na~{\sc i}~D. In this figure, the very narrow absorption features near the rest wavelengths of the D lines are due to interstellar Na~{\sc i} in the Milky Way. The broader, redshifted absorption features, which reach almost zero intensity at the core, are presumably due to interstellar material in the spiral arm of M51, which is seen in projection against NGC~5195. The M51 components are superimposed onto very broad stellar Na~{\sc i}~D lines, which are further redshifted and presumably arise from red giant stars in the nucleus of NGC~5195.

In addition to Na~{\sc i}~D, we clearly detect the interstellar K~{\sc i} and Ca~{\sc ii} doublets toward NGC~5195 near the velocity expected for gas associated with M51 (i.e., near $v_{\mathrm{LSR}}\approx430$~km~s$^{-1}$; see Section 3.1). The narrow Galactic component seen in Na~{\sc i}~D absorption (Figure~1), which has a velocity of $v_{\mathrm{LSR}}\approx$~$-$$30$~km~s$^{-1}$, is also detected in K~{\sc i} and absorption near this velocity can be seen in Ca~{\sc ii}~K, as well. However, the main focus of this investigation will be on the interstellar absorption features that we associate with the spiral arm of M51. We detect a strong line of interstellar CH$^+$~$\lambda4232$ toward NGC~5195 near the expected M51 velocity, but do not see evidence for CH~$\lambda4300$. (The non-detection of CH is not that unexpected given that the S/N in this region is only $\sim$30.) We do not detect any interstellar absorption features toward the nucleus of M51 other than a narrow Na~{\sc i}~D component associated with the Milky Way, which is weaker than the corresponding feature toward NGC~5195.

Eight DIBs are detected toward NGC~5195 at the same velocity as the other interstellar species that we presume to be associated with M51. The detected DIBs are those with rest wavelengths of 5780.5, 5797.1, 5849.8, 6196.0, 6203.6, 6376.1, 6379.3, and 6613.6~\AA{} (where we have adopted the convention of Friedman et al.~(2011) of quoting DIB wavelengths to one decimal place). Searches for other prominent DIBs in the NGC~5195 spectrum, such as the others studied by Friedman et al.~(2011) or those detected by Sollerman et al.~(2005) toward SN~2001el in NGC~1448, were unsuccessful. However, relatively weak and/or significantly broadened DIBs would be difficult to positively identify in our moderate S/N data, given the shape of the underlying continuum, which comprises many broad and often blended stellar absorption features. The DIBs that we do identify are generally much narrower than the underlying stellar features, as illustrated in Figure~2, which shows a portion of the NGC~5195 spectrum in the vicinity of the (redshifted) $\lambda5780.5$ and $\lambda5797.1$ DIBs. Also plotted in Figure~2 is a smoothed spectrum of the prototypical red giant star Arcturus (obtained from the atlas spectrum of Arcturus published in electronic form by Hinkle et al.~2000), which provides a template for the underlying stellar absorption features in the NGC~5195 spectrum. The close correspondence between the smoothed Arcturus spectrum and our adopted continuum fit for the region surrounding the $\lambda5780.5$ and $\lambda5797.1$ DIBs indicates that the continuum has been reasonably well fit and that any contamination of the DIB absorption by stellar lines is minimal. Similar comparisons for the regions surrounding the other detected DIBs indicate that the continua in those regions have also been fit appropriately.

In Figure~3, we present continuum-normalized spectra for the interstellar atomic and molecular species observed in the direction of NGC~5195. We also show (in the top panel of Figure~3) the H~{\sc i}~21~cm emission profile of M51 at the location of NGC~5195 provided by The H~{\sc i} Nearby Galaxy Survey (THINGS; Walter et al.~2008). This spectrum was extracted from the THINGS data cube\footnote{Available from the THINGS website at: http://www.mpia-hd.mpg.de/THINGS/} for M51 using the average of the four pixels nearest to the coordinates of NGC~5195. Figure~4 presents the absorption profiles for the eight DIBs identified in our co-added spectrum of NGC~5195. In Figure~5, we show continuum-normalized emission profiles for the [N~{\sc ii}]~$\lambda\lambda6548,6583$, H$\alpha$, and [S~{\sc ii}]~$\lambda\lambda6716,6730$ emission lines, which are prominent in the NGC~5195 spectrum. In Figure~6, we present the absorption profiles of the Ca~{\sc ii} infrared triplet, which (presumably) traces the red giant population in NGC~5195.

\section{ANALYSIS}
\subsection{Radial Velocities and Line Widths}
Radial velocities and line widths for the prominent emission lines and for several of the stellar absorption lines seen in the spectrum of NGC~5195 and of the nucleus of M51 were determined by fitting Gaussian functions to the observed features within IRAF (see Tables~1 and 2). For both galaxies, we note a relatively small, yet statistically significant, offset between the radial velocities of the emission lines and those of the stellar absorption features. For NGC~5195, the mean centroid velocity of the stellar absorption lines listed in Table~1 is $616\pm9$~km~s$^{-1}$, while the mean velocity of the emission lines is $570\pm7$~km~s$^{-1}$. (Errors for mean values reported in this paper are 1$\sigma$ standard deviations.) The overall mean for the emission lines and stellar absorption lines together is $588\pm25$~km~s$^{-1}$, which is in good agreement with the velocity listed for NGC~5195 in the Updated Zwicky Catalog (UZC): $v_{\mathrm{LSR}}=584\pm43$~km~s$^{-1}$ (Falco et al.~1999, applying a heliocentric-to-LSR correction of 12~km~s$^{-1}$). In the case of M51, the emission lines listed in Table~2 show a mean velocity of $502\pm7$~km~s$^{-1}$, while the stellar absorption lines have a mean of $477\pm7$~km~s$^{-1}$. The overall mean is $495\pm14$~km~s$^{-1}$, which (likewise) is consistent with the UZC value for M51: $v_{\mathrm{LSR}}=477\pm61$~km~s$^{-1}$.

The central velocity of the global H~{\sc i}~21~cm emission profile of M51 from THINGS is $v_{\mathrm{LSR}}=468$~km~s$^{-1}$ (Walter et al.~2008, applying the same heliocentric-to-LSR correction as above). At the position of NGC~5195, however, the velocity field (moment 1 map) of M51 provided by THINGS indicates an intensity-weighted mean H~{\sc i} velocity of $v_{\mathrm{LSR}}=434$~km~s$^{-1}$. The mean column-density weighted velocity of the interstellar atomic and molecular absorption lines seen in the direction of NGC~5195 is $411\pm4$~km~s$^{-1}$, and (similarly) the mean velocity exhibited by the DIBs is $411\pm5$~km~s$^{-1}$ (see Tables~3 and 4). Since these velocities are close to the mean H~{\sc i} velocity from THINGS at the position of NGC~5195, we can unambiguously associate the interstellar features with gas in M51 rather than in its companion. We do note, however, the slight difference between the H~{\sc i} velocity and the velocities of the interstellar absorption features toward NGC~5195, which is clearly discernible in Figure~3. We discuss this apparent discrepancy further in Section 4.

From the widths of the stellar absorption lines and the emission lines listed in Tables 1 and 2, we can derive estimates for the stellar velocity dispersion ($\sigma_*$) and the gaseous velocity dispersion ($\sigma_g$) in NGC~5195 and in the nucleus of M51. For a Gaussian line profile, the full width at half maximum (FWHM) is related to the velocity dispersion according to FWHM~=~$2\sqrt{2\ln2}\sigma$~$\approx$~$2.35\sigma$. The values of $\sigma$ derived using this relation are listed in the last columns of Tables~1 and 2. In converting the observed FWHM to $\sigma$, we have taken into account both the instrumental broadening (which, for our observations, is almost negligible), and the intrinsic widths of the stellar absorption lines. Estimates for the intrinsic widths of the stellar lines were obtained from the (unsmoothed) atlas spectrum of Arcturus (Hinkle et al.~2000), which we assume to be typical of the red giants in the nuclei of these galaxies. For NGC~5195, the stellar absorption lines yield a mean velocity dispersion of $\sigma_*=112\pm13$~km~s$^{-1}$, while the emission lines show a mean of $\sigma_g=70\pm10$~km~s$^{-1}$. For the nucleus of M51, the mean stellar velocity dispersion is $\sigma_*=91\pm16$~km~s$^{-1}$, while the mean gaseous velocity dispersion is $\sigma_g=72\pm13$~km~s$^{-1}$. Our estimates of $\sigma_*$ for these galaxies are consistent with the values adopted by Ho et al.~(2009), who give $\sigma_*=124.8\pm8.1$~km~s$^{-1}$ for NGC~5195 and $\sigma_*=96.0\pm8.7$~km~s$^{-1}$ for M51. However, our results also suggest that $\sigma_*>\sigma_g$ for both NGC~5195 and M51 (although the difference for M51 may not be significant). When considered together, the differences between the radial velocities and velocity dispersions inferred from the stellar absorption lines and those derived from the emission features may be indicative of real differences in the kinematics of the stars and the ionized gas in the central regions of these galaxies.

\subsection{Equivalent Widths and Column Densities of the Interstellar Features}
The equivalent widths ($W_{\lambda}$) of the interstellar atomic and molecular absorption lines observed in the direction of NGC~5195 (that are associated with interstellar gas in M51) were computed by direct integration over the absorption profiles (see Table~3). The limits of integration were set according to the locations where the optical depth reached zero. For doublets, the limits were determined based on the stronger line and applied in an identical way to the weaker line. Uncertainties in the integrated equivalent widths include contributions from photon noise and continuum placement. The photon noise uncertainties were derived using the formula given in Jenkins et al.~(1973), which can be written $\sigma_{W_{\lambda}}=(\Delta\lambda)(N_{\mathrm{p}})^{1/2}(\mathrm{RMS})$, where $\Delta\lambda$ is the difference in wavelength between adjacent pixels, $N_{\mathrm{p}}$ is the number of pixels sampling the absorption profile, and RMS is the root mean square of the noise in the continuum. The DIB equivalent widths and uncertainties were computed in the same manner (see Table~4). As an aid in determining the limits of integration for the DIB equivalent widths, we compared the profiles of the DIBs detected toward NGC~5195 with those seen toward Galactic stars (e.g., Friedman et al.~2011). The adopted limits are explicitly shown in Figure~4.

Column densities for the atomic and molecular species were determined by means of the apparent optical depth (AOD) method, which is a useful approach when it is suspected that the absorption profiles may be affected by unresolved saturation (see Savage \& Sembach 1991). While there is evidence for component structure in the atomic and molecular absorption profiles, in the form of a weak wing redward of the main absorption components (see Figure~3), the moderate resolution of our observations and the relatively low S/N achieved below 5000~\AA{} suggests that a detailed profile synthesis analysis would not be an ideal approach for these data. However, we can still recover measurements of total column densities through use of the AOD method. Apparent column densities ($N_a$) were derived by integrating the AOD profiles over the same intervals used to compute the equivalent widths. Values of $N_a$ for the individual lines detected toward NGC~5195 are listed in Table~3. Uncertainties in these quantities were determined through standard error propagation (see, e.g., Prochaska et al.~2001), and also include continuum placement uncertainties.

A comparison between the apparent column densities of the two K~{\sc i} transitions indicates that these features may be very mildly saturated. The difference in $N_a$ for the two lines suggests that a correction of +0.04 dex (applied to the weaker line) is appropriate, according to the correction scheme devised by Savage \& Sembach (1991). This correction has been used to derive the adopted column density ($N_{\mathrm{adopt}}$) for K~{\sc i} listed in Table~3. A similar comparison for the two Ca~{\sc ii} transitions indicates that a correction of +0.12 dex is needed, and this has been applied to the weaker line to derive the adopted Ca~{\sc ii} column density. The interstellar Na~{\sc i} lines toward NGC~5195 (that are associated with M51) are very heavily saturated. The difference in $N_a$ for the two transitions suggests that a correction of at least +0.49 dex is necessary. Yet, while this correction has been applied to the weaker line, we still regard the resulting Na~{\sc i} column density as a lower limit. No saturation correction is needed in the case of CH$^+$ $\lambda4232$. A tentative detection of the weaker CH$^+$ $\lambda3957$ line yields a CH$^+$ column density consistent with that given in Table~3.

\section{DISCUSSION}
The column densities of the interstellar atomic and molecular species and the equivalent widths of the DIBs detected toward NGC~5195 provide a means to estimate some of the basic properties of the intervening interstellar material in the spiral arm of M51. Several well-defined correlations are known to exist between various interstellar quantities for sight lines in the local Galactic ISM (e.g., Welty et al.~2014), and, here, we explore what these correlations would imply for the M51 material toward NGC~5195. Specifically, we consider the known correlations between the equivalent width of the $\lambda5780.5$ DIB and the column density of atomic hydrogen $N$(H~{\sc i}) (e.g., Herbig  1993; Welty et al.~2006; Friedman et al.~2011), between the K~{\sc i} column density and the total hydrogen column density $N$(H$_{\mathrm{tot}}$)~=~$N$(H~{\sc i})~+~2$N$(H$_2$) and/or the H$_2$ column density (e.g., Welty \& Hobbs 2001), between the column densities of CH and H$_2$ (e.g., Welty et al.~2006; Sheffer et al.~2008), and between the equivalent widths of various DIBs and $E(\bv)$ (e.g., Friedman et al.~2011; Vos et al.~2011).

To derive specific estimates for $N$(H~{\sc i}), $N$(H$_2$), $N$(H$_{\mathrm{tot}}$), and $E(\bv)$ from the quantities we measure toward NGC~5195, we use the recently updated Galactic relationships involving these quantities discussed in Welty et al.~(2014) and Ritchey et al.~(2015). Our measurement of $W_{\lambda}$(5780.5)~=~$71\pm10$~m\AA{} implies an H~{\sc i} column density of log~$N$(H~{\sc i})~=~$20.75\pm0.12$. This value (perhaps fortuitously) is identical to that obtained from the THINGS integrated H~{\sc i} (moment 0) map of M51 at the position of NGC~5195; the average of the four pixels nearest to the coordinates of NGC~5195 in the THINGS map indicates a column density of log~$N$(H~{\sc i})~=~$20.75\pm0.06$. In contrast, the total hydrogen column density implied by the column density of K~{\sc i} toward NGC~5195 is significantly higher; our measurement of log~$N$(K~{\sc i})~=~$12.11\pm0.01$ implies log~$N$(H$_{\mathrm{tot}}$)~=~$21.43\pm0.15$. Taken together, these estimates for $N$(H~{\sc i}) and $N$(H$_{\mathrm{tot}}$) would imply an H$_2$ column density of log~$N$(H$_2$)~=~$21.03\pm0.18$, and a molecular fraction of $f$(H$_2$)~=~2$N$(H$_2$)/$N$(H$_{\mathrm{tot}}$)~=~$0.79\pm0.18$. A somewhat lower H$_2$ column density is obtained if we estimate $N$(H$_2$) directly from $N$(K~{\sc i}); in this case, we find log~$N$(H$_2$)~=~$20.73\pm0.17$, which, when combined with our estimate for $N$(H~{\sc i}), would give log~$N$(H$_{\mathrm{tot}}$)~=~$21.21\pm0.12$ and $f$(H$_2$)~=~$0.65\pm0.19$. We note that a relatively high value of $N$(H$_2$) is not ruled out by the non-detection of CH, as our 3$\sigma$ upper limit of log~$N$(CH)~$\lesssim13.6$ implies log~$N$(H$_2$)~$\lesssim21.0$. However, this upper limit does tend to favor the lower H$_2$ column density estimated directly from $N$(K~{\sc i}).

There is reason to suspect that the total hydrogen column density predicted from $N$(K~{\sc i}) toward NGC~5195 may be an overestimate. Deviations from the usual Galactic relationship between $N$(K~{\sc i}) and $N$(H$_{\mathrm{tot}}$) have been noted for particular regions in the local Galactic ISM and for certain extragalactic sight lines. Many sight lines in the Sco-Oph and Orion Trapezium regions, for example, exhibit lower values of $N$(K~{\sc i}) than would be expected based on their total hydrogen column densities, most likely as a result of enhanced radiation fields in those environments (Welty \& Hobbs 2001). Low values of $N$(K~{\sc i}), relative to $N$(H$_{\mathrm{tot}}$), are also observed for sight lines in the LMC and SMC; in these cases, the lower values probably reflect the combined effects of stronger radiation fields and lower overall metallicities (Welty \& Hobbs 2001; Welty 2014). Differences in metallicity and/or in the strength of the radiation field for the M51 material toward NGC~5195 could therefore (potentially) be affecting our estimate of $N$(H$_{\mathrm{tot}}$) from $N$(K~{\sc i}). The H$_2$ column density predicted from $N$(K~{\sc i}) should not be as affected, however, since there seems to be less regional variation in the $N$(K~{\sc i}) versus $N$(H$_2$) relation, probably because both K~{\sc i} and H$_2$ can be affected by the local radiation field (D.~Welty 2014, private communication). We note also that a solar metallicity for the M51 material is suggested by the fact that the equivalent width of the $\lambda5780.5$ DIB, when compared with the H~{\sc i} column density derived from THINGS data, is consistent with the Galactic relationship between $W_{\lambda}$(5780.5) and $N$(H~{\sc i}) (e.g., Welty et al.~2006; Friedman et al.~2011). In the discussion that follows, we will adopt the value of $N$(H$_2$) estimated directly from $N$(K~{\sc i}) (i.e., log~$N$(H$_2$)~$\approx$~$20.73$), along with the corresponding values for $N$(H$_{\mathrm{tot}}$) and $f$(H$_2$) assuming log~$N$(H~{\sc i})~$\approx$~$20.75$.

An estimate for the interstellar reddening associated with the M51 material toward NGC~5195 can be obtained from the strengths of the various DIBs observed in that direction. Again, we use the updated correlations between $W_{\lambda}$(DIB) and $E(\bv)$ reported in Welty et al.~(2014), who studied all of the DIBs we detect toward NGC~5195 except for $\lambda5849.8$ and $\lambda6376.1$. The measured equivalent widths of the other six DIBs yield values of $E(\bv)$ in the range 0.13--0.60 mag, with a mean of $0.35\pm0.20$ mag (see Table~4). However, in Galactic sight lines with high molecular fractions, many of the typically stronger, well-studied ``standard'' DIBs (such as those detected toward NGC~5195) tend to be weaker than expected compared to the general trends with $N$(H$_{\mathrm{tot}}$) and $E(\bv)$, but are not weaker compared to trends with $N$(H~{\sc i}) (Welty et al.~2014), suggesting that the standard DIBs trace primarily atomic gas. This may help to explain the weakness of the $\lambda5780.5$, $\lambda6196.0$, $\lambda6379.3$, and $\lambda6613.6$ DIBs toward NGC~5195, which are weaker by 75\%, 70\%, 52\%, and 44\%, respectively, compared to the strengths predicted from the amount of K~{\sc i} present (see Table~4). The strengths of the $\lambda5797.1$ and $\lambda6203.6$ DIBs toward NGC~5195 are consistent with the predictions based on $N$(K~{\sc i}), considering the uncertainties. Still, the DIBs toward NGC~5195 are, on average, about 40\% weaker than predicted, meaning that the $E(\bv)$ values derived from the DIB equivalent widths are likely to be underestimates.

We can refine our estimate for $E(\bv)$ by considering the relationships that exist between the residuals of the observed values of $W_{\lambda}$(DIB) with respect to the mean Galactic trends with both $N$(K~{\sc i}) and $E(\bv)$ (Welty et al.~2014).\footnote{For a given DIB, the ``residual'' is here defined as the difference between (the logarithm of) the measured equivalent width and (the logarithm of) the equivalent width predicted by the appropriate Galactic relationship involving either $N$(K~{\sc i}) or $E(\bv)$.} All of the DIBs studied by Welty et al.~(2014) exhibit positive slopes in (logarithmic) plots of residual with respect to $E(\bv)$ versus residual with respect to $N$(K~{\sc i}), indicating that when a given DIB is weak relative to $N$(K~{\sc i}), it will also be weak relative to $E(\bv)$. Use of the relationships for the residuals provided by Welty et al.~(2014) for the DIBs detected toward NGC~5195 yields the values of $E(\bv)$ given in the last column of Table~4. Compared to the original predicted values, these adjusted values exhibit a narrower range (0.35--0.55 mag), along with a higher mean value and smaller standard deviation ($0.46\pm0.09$ mag), and should be more representative of the reddening toward NGC~5195 that is associated with intervening interstellar material in M51.

Our estimates of log~$N$(H$_{\mathrm{tot}}$)~$\approx$~$21.21$ and $E(\bv)$~$\approx$~$0.46$ mag for the M51 material toward NGC~5195 yield a gas-to-dust ratio of $N$(H$_{\mathrm{tot}}$)/$E(\bv)$~$\approx$~$3.6\times10^{21}$~cm$^{-2}$~mag$^{-1}$, somewhat lower than the standard Galactic gas-to-dust ratio of $5.6$--$5.9\times10^{21}$~cm$^{-2}$~mag$^{-1}$ (Bohlin et al.~1978; Rachford et al.~2009; Welty et al.~2012). For the atomic gas alone, we find a gas-to-dust ratio of $N$(H~{\sc i})/$E(\bv)$~$\approx$~$1.2\times10^{21}$~cm$^{-2}$~mag$^{-1}$, which (likewise) is below the Galactic average of $4.4$--$5.2\times10^{21}$~cm$^{-2}$~mag$^{-1}$, but still within the tail of the distribution for Galactic sight lines (Welty et al.~2012). The corresponding ratio for the molecular gas, $N$(H$_2$)/$E(\bv)$~$\approx$~$1.2\times10^{21}$~cm$^{-2}$~mag$^{-1}$, is somewhat higher than the average Galactic ratio, but well within the Galactic distribution. The line of sight to NGC~5195 therefore seems to be probing interstellar material in M51 with a fairly high molecular content, but with a somewhat lower overall gas-to-dust ratio compared to that most commonly seen for sight lines in the Milky Way.

The H~{\sc i} column density that we derive for the M51 material toward NGC~5195 (i.e., log~$N$(H~{\sc i})~$\approx$~$20.75$) should be secure since we obtain the same result from both emission and absorption features. Normally, when one compares a column density deduced from 21~cm emission to that obtained from an absorption-line tracer, there is some ambiguity related to the exact location of the background source; it is not usually known how much of the emitting gas is being probed by the source used for the absorption-line measurements. In this case, however, we are confident that the nucleus of NGC~5195 probes the entire line of sight through the spiral arm of M51. Furthermore, there should not be any discrepancies between the H~{\sc i} column densities derived from the optical absorption and 21~cm emission features due to structure transverse to the line of sight since the beam size for THINGS ($\sim$$6\arcsec$; Walter et al.~2008) is comparable to the size of the slit used for the echelle observations ($\sim$$3\farcs2$).

It is therefore somewhat surprising that the H~{\sc i}~21~cm emission profile does not show better agreement in velocity space with the interstellar absorption features observed in the direction of NGC~5195 (Figure~3). The intensity-weighted mean H~{\sc i} velocity is $v_{\mathrm{LSR}}=434$~km~s$^{-1}$, while the atomic and molecular absorption lines show a mean column-density weighted velocity of $v_{\mathrm{LSR}}=411$~km~s$^{-1}$. Figure~3 clearly demonstrates that the absorption peak is shifted to lower velocities compared to the peak in H~{\sc i} emission. Moreover, the extent of the H~{\sc i} emission seems to coincide better with the weaker absorption components seen to the red of the main components in the Na~{\sc i}, K~{\sc i}, Ca~{\sc ii}, and CH$^+$ profiles. Since these latter species trace both atomic and molecular gas, it seems likely that the material ``missing'' from the H~{\sc i} profile at the velocity of the main interstellar absorption components is mostly molecular.

There are other indications that the line of sight to NGC~5195 is probing relatively dense gas in M51. Our lower limit on the Na~{\sc i} column density of log~$N$(Na~{\sc i})~$\gtrsim14.0$, combined with our determination of log~$N$(Ca~{\sc ii})~=~12.96, yields $N$(Na~{\sc i})/$N$(Ca~{\sc ii})~$\gtrsim11$. High $N$(Na~{\sc i})/$N$(Ca~{\sc ii}) ratios such as this are usually associated with cold, dense quiescent clouds, where Ca is heavily depleted within interstellar grains. While the $N$(Na~{\sc i})/$N$(Ca~{\sc ii}) ratio can also be affected by variations in the ionization conditions (which can alter the abundances of both Na~{\sc i} and Ca~{\sc ii}), the dominant effect in this case is probably the depletion of Ca. At log~$N$(Na~{\sc i})~$\gtrsim14.0$, Ca~{\sc ii} is likely to be the dominant ionization state of Ca in this material (e.g., Welty et al.~1996). We can therefore derive the Ca depletion factor directly from the Ca~{\sc ii} column density by combining this value with our estimate for $N$(H$_{\mathrm{tot}}$). Assuming that the metallicity of the M51 material is roughly solar, and adopting the solar Ca abundance from Lodders~(2003), we find [Ca/H]~=~log~$N$(Ca~{\sc ii})~$-$~log~$N$(H$_{\mathrm{tot}}$)~$-$~log~(Ca/H)$_{\sun}$~=~$-$$2.66\pm0.15$, which is comparable to some of the largest depletions seen for Ti in the local Galactic ISM (e.g., Jenkins~2009; Welty \& Crowther 2010).

We have already remarked on the general weakness of the DIBs toward NGC~5195, which may be an indication that the gas being probed has a high molecular content. We further note that the ratio of the equivalent widths of the $\lambda5797.1$ and $\lambda5780.5$ DIBs is quite high; we find $W_{\lambda}$(5797.1)/$W_{\lambda}$(5780.5)~=~$1.55\pm0.26$, which, to our knowledge, is among the highest values ever measured for this ratio (e.g., Vos et al.~2011; Welty et al.~2014). Such a high value for $W_{\lambda}$(5797.1)/$W_{\lambda}$(5780.5) may suggest either that the material is immersed in a very weak ambient radiation field, or else that the environment is significantly shielded from UV radiation (Vos et al.~2011). A low-UV environment for the M51 material could potentially explain why the total hydrogen column density predicted directly from $N$(K~{\sc i}) seems to be somewhat high (compared to our adopted value). A factor-of-two enhancement in $N$(K~{\sc i}), due to a weak radiation field, would be sufficient to explain the difference in the predictions for $N$(H$_{\mathrm{tot}}$). As high $W_{\lambda}$(5797.1)/$W_{\lambda}$(5780.5) ratios also appear to be associated with high molecular fractions (Weselak et al.~2004), the high value of $W_{\lambda}$(5797.1)/$W_{\lambda}$(5780.5) that we find for the M51 material toward NGC~5195 is consistent (qualitatively) with our estimate of $f$(H$_2$)~$\approx$~0.65.

One complication to this picture is that the $N$(CH$^+$)/$N$(CH) ratio toward NGC~5195 seems also to be fairly high; our upper limit on the CH column density, combined with our measurement of log~$N$(CH$^+$)~=~13.91, gives $N$(CH$^+$)/$N$(CH)~$\gtrsim2.3$, while, for relatively dense and/or significantly shielded gas with a high molecular fraction, this ratio would be expected to be $\lesssim1.0$ (e.g., Pan et al.~2005; Welty et al.~2006; Ritchey et al.~2006; Sheffer et al.~2008). A similar result was found for the ISM of M82 along the line of sight to SN~2014J (Welty et al.~2014; Ritchey et al.~2015). The material in that direction was shown to have an unusual combination of a high $W_{\lambda}$(5797.1)/$W_{\lambda}$(5780.5) ratio and a high $N$(CH$^+$)/$N$(CH) ratio, compared to sight lines in the Milky Way, in the LMC and SMC, and in other supernova host galaxies. Compared to the M82 gas toward SN~2014J, the material in M51 toward NGC~5195 may have a somewhat lower $N$(CH$^+$)/$N$(CH) ratio and has a higher $W_{\lambda}$(5797.1)/$W_{\lambda}$(5780.5) ratio, but the M51 results still seem to be offset from the results for other sight lines (see, e.g., Figure~5 in Welty et al.~2014).

The high CH$^+$ abundances in the M51 gas toward NGC~5195 and the M82 gas toward SN~2014J suggest that non-thermal chemical processes are highly active in these galaxies (e.g., Zsarg{\'o} \& Federman 2003). The formation of CH$^+$ occurs through the endothermic reaction C$^+$~+~H$_2$~$\rightarrow$~CH$^+$~+~H, which cannot proceed at the low temperatures of diffuse molecular clouds (due to its activation energy of $\Delta E/k_{\mathrm{B}}=4640$~K). Thus, the large abundances of CH$^+$ observed in the Galactic ISM have traditionally been difficult to explain. Joulain et al.~(1998) proposed that intermittent bursts of turbulent dissipation would be sufficient to heat the gas locally so as to account for the observed CH$^+$ abundances. Their model of turbulent dissipation regions (TDRs) has been further developed by Godard et al.~(2009, 2014), who find that the CH$^+$ abundance is directly proportional to the average turbulent dissipation rate. Tubulence is enhanced in the ISM of interacting galaxies (e.g., Irwin 1994; Elmegreen et al.~1995), and both M51 and M82 are known to be interacting with their neighbors (M51 with NGC~5195 and M82 with M81). A possible explanation, therefore, of the high CH$^+$ abundances in M51 and M82 is that these abundances are linked to enhancements in interstellar turbulence resulting from galaxy-galaxy interactions. Such enhancements could be directly related to the gravitational encounters occurring between the interacting galaxies or could be due to subsequent increases in the star formation and supernova rates triggered by those interactions.

\section{CONCLUDING REMARKS}

We have discussed moderately-high resolution echelle observations of the nucleus of NGC~5195, the line of sight to which probes interstellar gas associated with the northeastern spiral arm of M51. Absorption from Na~{\sc i}, K~{\sc i}, Ca~{\sc ii}, and CH$^+$ is detected at a velocity close to that exhibited by H~{\sc i}~21~cm emission from M51 at the position of NGC~5195. However, the H~{\sc i} emission profile (provided by THINGS) peaks at a slightly higher velocity, compared to the atomic and molecular absorption profiles, and coincides better with weaker absorption components seen to the red of the main components. Eight DIBs are detected at velocities similar to those of the atomic and molecular species seen in absorption. The H~{\sc i} column density implied by the equivalent width of the $\lambda5780.5$ DIB is consistent with that obtained from the THINGS integrated H~{\sc i} map of M51 at the location of NGC~5195. Intriguingly, the H$_2$ column density predicted from the observed column density of K~{\sc i} is comparable to $N$(H~{\sc i}), suggesting a high molecular fraction ($\sim$0.65) for the M51 gas toward NGC~5195. The DIBs toward NGC~5195 are, on average, about 40\% weaker than would be expected based on the K~{\sc i} column density, which is a further indication that the gas in this direction has a high molecular content. After correcting for the overall weakness of the DIBs, we estimate the reddening toward NGC~5195 to be $E(\bv)$~$\approx$~0.46 mag. The M51 material is characterized also by a high $N$(Na~{\sc i})/$N$(Ca~{\sc ii}) ratio ($\gtrsim11$), indicative of a high degree of Ca depletion, and a high $W_{\lambda}$(5797.1)/$W_{\lambda}$(5780.5) ratio ($\sim$1.6), suggestive of either a very weak ambient radiation field or a significantly shielded environment. A high $N$(CH$^+$)/$N$(CH) ratio ($\gtrsim2.3$) for the gas in this direction may be the result of enhanced turbulence due to interactions between M51 and its companion.

While there have been other absorption-line studies of moderately dense gas in nearby galaxies, typically using a supernova as the continuum source (e.g., Cox \& Patat 2008, 2014), an important aspect of this investigation is that the background source is not a supernova, and hence will not fade with time. This opens up the possibility of further high-resolution observations, which could explore in more detail the physical and chemical properties of the interstellar gas in M51 toward NGC~5195. For example, ground-based observations with higher S/N and higher resolution than those presented here could target absorption from Li~{\sc i}, and could potentially yield the first determination of the Li isotope ratio in a galaxy beyond the Local Group. A more comprehensive examination of the DIBs in this direction could yield important information concerning the behavior of the DIBs in an environment different from that typically encountered in the local Galactic ISM. Finally, observations of NGC~5195 with the Cosmic Origins Spectrograph onboard the \emph{Hubble Space Telescope} would provide access to important UV absorption-line tracers, which could be used to study the abundances and physical conditions in the M51 material in much greater detail than can be done from the ground. Such information would be particularly useful in more confidently identifying the mechanism responsible for the high CH$^+$ abundance, and could ultimately provide valuable constraints for models of CH$^+$ formation in the ISM of our Galaxy and beyond.

\acknowledgements
We thank Dan Welty for offering many useful comments and suggestions on this work. Support for this research was provided by the Kenilworth Fund of the New York Community Trust. The results presented here are based on observations obtained with the Apache Point Observatory 3.5-m telescope, which is owned and operated by the Astrophysical Research Consortium.

\begin{deluxetable}{lccccc}
\tablecolumns{6}
\tablewidth{0pc}
\tabletypesize{\small}
\tablecaption{Absorption and Emission Features in NGC~5195\label{t1}}
\tablehead{
\colhead{Species} & \colhead{$\lambda$} & \colhead{$\lambda_{\mathrm{obs}}$} & \colhead{$v_{\mathrm{LSR}}$} & \colhead{FWHM} & \colhead{$\sigma$} \\
\colhead{} & \colhead{(\AA)} & \colhead{(\AA)} & \colhead{(km s$^{-1}$)} & \colhead{(km s$^{-1}$)} & \colhead{(km s$^{-1}$)} \\
}
\startdata
\multicolumn{6}{c}{Stellar Absorption Lines} \\
\hline
Mg~{\sc i} & 5183.60 & 5194.07 & $+606$ & 252 & 99 \\
Ca~{\sc ii} & 8498.02 & 8515.80 & $+627$ & 279 & 118 \\
 & 8542.09 & 8559.58 & $+614$ & 309 & 128 \\
 & 8662.14 & 8679.95 & $+616$ & 249 & 103 \\
\hline
\multicolumn{6}{c}{Emission Lines} \\
\hline
H$\alpha$ & 6562.82 & 6574.94 & $+563$ & 157 & 66 \\
\lbrack{}O~{\sc i}\rbrack{} & 6300.30 & 6312.45 & $+578$ & 125 & 53 \\
\lbrack{}N~{\sc ii}\rbrack{} & 6548.05 & 6560.65 & $+577$ & 171 & 72 \\
 & 6583.45 & 6595.78 & $+561$ & 194 & 82 \\
\lbrack{}S~{\sc ii}\rbrack{} & 6716.44 & 6729.16 & $+568$ & 169 & 72 \\
 & 6730.82 & 6743.66 & $+572$ & 170 & 72 \\
\enddata
\end{deluxetable}

\begin{deluxetable}{lccccc}
\tablecolumns{6}
\tablewidth{0pc}
\tabletypesize{\small}
\tablecaption{Absorption and Emission Features in the Nucleus of M51\label{t2}}
\tablehead{
\colhead{Species} & \colhead{$\lambda$} & \colhead{$\lambda_{\mathrm{obs}}$} & \colhead{$v_{\mathrm{LSR}}$} & \colhead{FWHM} & \colhead{$\sigma$} \\
\colhead{} & \colhead{(\AA)} & \colhead{(\AA)} & \colhead{(km s$^{-1}$)} & \colhead{(km s$^{-1}$)} & \colhead{(km s$^{-1}$)} \\
}
\startdata
\multicolumn{6}{c}{Stellar Absorption Lines} \\
\hline
Mg~{\sc i} & 5183.60 & 5191.78 & $+473$ & 206 & 78 \\
Ca~{\sc ii} & 8498.02 & 8511.71 & $+483$ & 233 & 98 \\
 & 8542.09 & 8555.89 & $+484$ & 268 & 110 \\
 & 8662.14 & 8675.70 & $+469$ & 190 & 77 \\
\hline
\multicolumn{6}{c}{Emission Lines} \\
\hline
H$\alpha$ & 6562.82 & 6573.62 & $+502$ & 182 & 77 \\
H$\beta$ & 4861.33 & 4869.65 & $+513$ & 108 & 46 \\
\lbrack{}O~{\sc i}\rbrack{} & 6300.30 & 6310.64 & $+492$ & 138 & 59 \\
\lbrack{}O~{\sc iii}\rbrack{} & 4958.91 & 4967.38 & $+512$ & 166 & 71 \\
\lbrack{}O~{\sc iii}\rbrack{} & 5006.84 & 5015.23 & $+502$ & 195 & 83 \\
\lbrack{}N~{\sc ii}\rbrack{} & 6548.05 & 6558.96 & $+500$ & 184 & 78 \\
 & 6583.45 & 6594.41 & $+499$ & 216 & 92 \\
\lbrack{}S~{\sc ii}\rbrack{} & 6716.44 & 6727.57 & $+497$ & 176 & 75 \\
 & 6730.82 & 6742.10 & $+503$ & 172 & 73 \\
\enddata
\end{deluxetable}

\begin{deluxetable}{lcccccc}
\tablecolumns{7}
\tablewidth{0pc}
\tabletypesize{\small}
\tablecaption{Interstellar Atomic and Molecular Absorption Lines toward NGC 5195\label{t3}}
\tablehead{
\colhead{Species} & \colhead{$\lambda$\tablenotemark{a}} & \colhead{log $f\lambda$\tablenotemark{a}} & \colhead{$\langle v_{\mathrm{LSR}} \rangle$\tablenotemark{b}} & \colhead{$W_{\lambda}$} & \colhead{log $N_a$} & \colhead{log $N_{\mathrm{adopt}}$\tablenotemark{c}} \\
\colhead{} & \colhead{(\AA)} & \colhead{} & \colhead{(km s$^{-1}$)} & \colhead{(m\AA)} & \colhead{} & \colhead{} \\
}
\startdata
Ca~{\sc ii} & 3933.661 & 3.392 & $+405$ & $371\pm39$ & $12.73\pm0.05$ & $12.96\pm0.08$ \\
 & 3968.467 & 3.092 & $+410$ & $242\pm43$ & $12.84\pm0.07$ & \ldots \\
Na~{\sc i} & 5889.951 & 3.577 & $+415$ & $1556\pm8$\phn\phn & $13.27\pm0.01$ & $\gtrsim14.0$ \\
 & 5895.924 & 3.276 & $+414$ & $1403\pm8$\phn\phn & $13.49\pm0.01$ & \ldots \\
K~{\sc i} & 7664.899 & 3.710 & $+413$ & $325\pm6$\phn & $12.03\pm0.01$ & $12.11\pm0.01$ \\
 & 7698.965 & 3.409 & $+413$ & $191\pm5$\phn & $12.07\pm0.01$ & \ldots \\
CH$^+$ & 4232.548 & 1.363 & $+410$ & \phn$67\pm16$ & $13.91\pm0.10$ & $13.91\pm0.10$ \\
CH & 4300.313 & 1.338 & \ldots & $\lesssim30$ & $\lesssim13.6$ & $\lesssim13.6$ \\
\enddata
\tablenotetext{a}{Wavelengths and $f$-values are from Morton~(2003) for atomic lines and Gredel et al.~(1993) for molecular lines.}
\tablenotetext{b}{Column-density weighted mean velocity: $\langle v_{\mathrm{LSR}} \rangle = \big( \sum N_a(v) \times v \big) / \big( \sum N_a(v) \big)$.}
\tablenotetext{c}{Adopted column density, including any correction for line saturation.}
\end{deluxetable}

\begin{deluxetable}{lccccccc}
\tablecolumns{8}
\tablewidth{0pc}
\tabletypesize{\small}
\tablecaption{Diffuse Interstellar Bands toward NGC 5195\label{t4}}
\tablehead{
\colhead{$\lambda$} & \colhead{$\lambda_{\mathrm{obs}}$\tablenotemark{a}} & \colhead{$v_{\mathrm{LSR}}$} & \colhead{FWHM\tablenotemark{a}} & \colhead{$W_{\lambda\mathrm{,meas}}$\tablenotemark{b}} & \colhead{$W_{\lambda\mathrm{,pred}}$\tablenotemark{c}} & \colhead{$E(\bv)_{\mathrm{pred}}$\tablenotemark{d}} & \colhead{$E(\bv)_{\mathrm{corr}}$\tablenotemark{e}} \\
\colhead{(\AA)} & \colhead{(\AA)} & \colhead{(km s$^{-1}$)} & \colhead{(\AA)} & \colhead{(m\AA)} & \colhead{(m\AA)} & \colhead{(mag)} & \colhead{(mag)} \\
}
\startdata
5780.5 & 5788.6 & $+418$ & 1.49 & \phn$71\pm10$ & 288 & 0.13 & 0.36 \\
5797.1 & 5805.1 & $+414$ & 1.07 & $110\pm10$ & 121 & 0.56 & 0.54 \\
5849.8 & 5857.8 & $+408$ & 1.02 & $42\pm6$ & \ldots & \ldots & \ldots \\
6196.0 & 6204.4 & $+408$ & 0.28 & $10\pm2$ & 33 & 0.16 & 0.35 \\
6203.6 & 6212.2 & $+416$ & 3.87 & $126\pm16$ & 110 & 0.60 & 0.55 \\
6376.1 & 6384.7 & $+406$ & 1.11 & $37\pm6$ & \ldots & \ldots & \ldots \\
6379.3 & 6388.0 & $+407$ & 0.56 & $26\pm5$ & 55 & 0.29 & 0.49 \\
6613.6 & 6622.7 & $+414$ & 1.16 & $75\pm6$ & 133 & 0.38 & 0.46 \\
\enddata
\tablenotetext{a}{Observed central wavelengths and band widths of the DIBs were determined according to the definitions given in Hobbs et al.~(2008).}
\tablenotetext{b}{Measured $W_{\lambda}$(DIB).}
\tablenotetext{c}{Predicted $W_{\lambda}$(DIB), assuming log $N$(K~{\sc i})~$\approx$~12.11 and using Galactic relationships from Welty et al.~(2014).}
\tablenotetext{d}{Predicted $E(\bv)$ from measured $W_{\lambda}$(DIB) and using Galactic relationships given in Welty et al.~(2014).}
\tablenotetext{e}{Predicted $E(\bv)$, adjusted for residuals, using Galactic relationships given in Welty et al.~(2014).}
\end{deluxetable}

\clearpage

\begin{figure}
\centering
\includegraphics[width=0.9\textwidth]{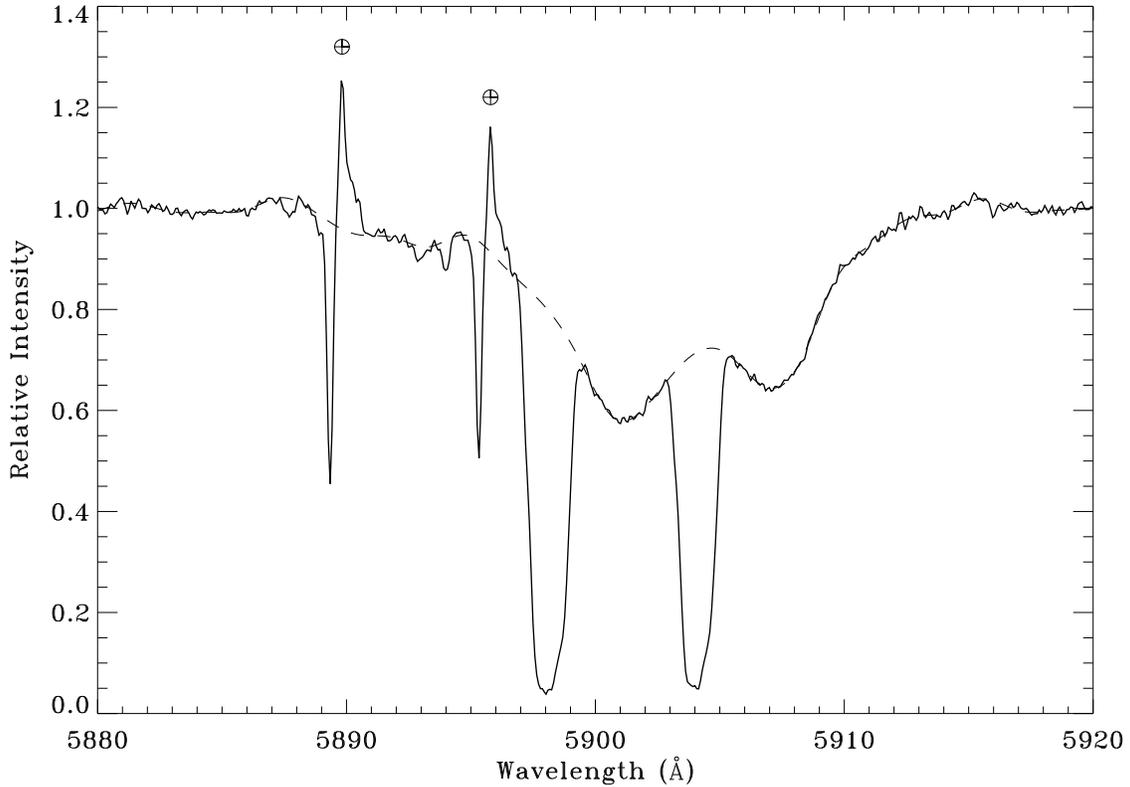}
\caption{Final co-added spectrum of NGC~5195 in the vicinity of Na~{\sc i}~D. Narrow Galactic Na~{\sc i} absorption components are seen near the rest wavelengths of the D lines. The broader, redshifted absorption features are presumably due to interstellar material in the spiral arm of M51. These features are superimposed onto very broad stellar Na~{\sc i}~D lines, which are further redshifted and presumably arise from red giant stars in the nucleus of NGC~5195. The dashed line shows the continuum fit adopted for the purpose of analyzing the interstellar features. Strong Na~{\sc i} night sky emission lines are identified with the symbol $\oplus$.}
\label{f1}
\end{figure}

\begin{figure}
\centering
\includegraphics[width=0.9\textwidth]{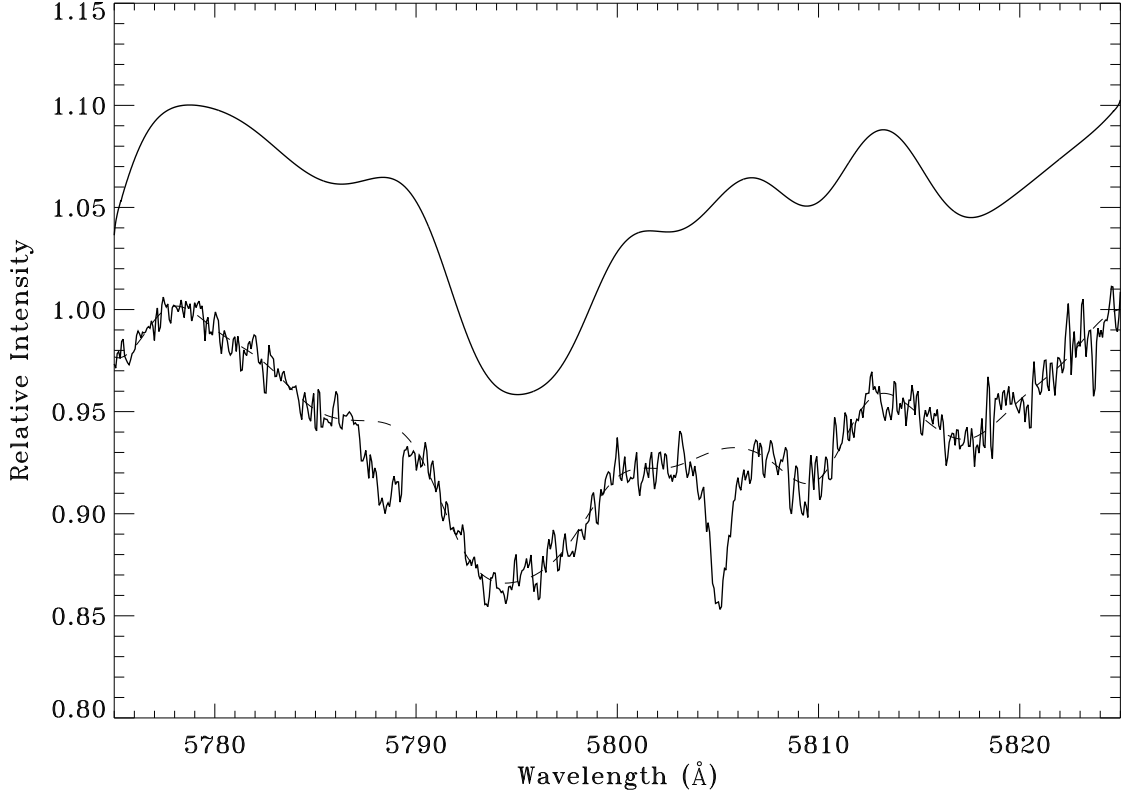}
\caption{Final co-added spectrum of NGC~5195 in the vicinity of the (redshifted) $\lambda5780.5$ and $\lambda5797.1$ DIBs (which can be seen at wavelengths of 5788.6 and 5805.1~\AA{}, respectively). The dashed line shows the adopted continuum obtained through a low-order polynomial fit to the observed spectrum (excluding the regions showing DIB absorption). Also shown (displaced by +0.1 in relative intensity units) is a smoothed spectrum of the prototypical red giant star Arcturus (Hinkle et al.~2000), which provides a template for the underlying stellar absorption features in the NGC~5195 spectrum. The Arcturus spectrum has been shifted by +616~km~s$^{-1}$ in accordance with the mean velocity exhibited by the stellar absorption lines in NGC~5195 (see Section 3.1).}
\label{f2}
\end{figure}

\begin{figure}
\centering
\includegraphics[width=0.49\textwidth]{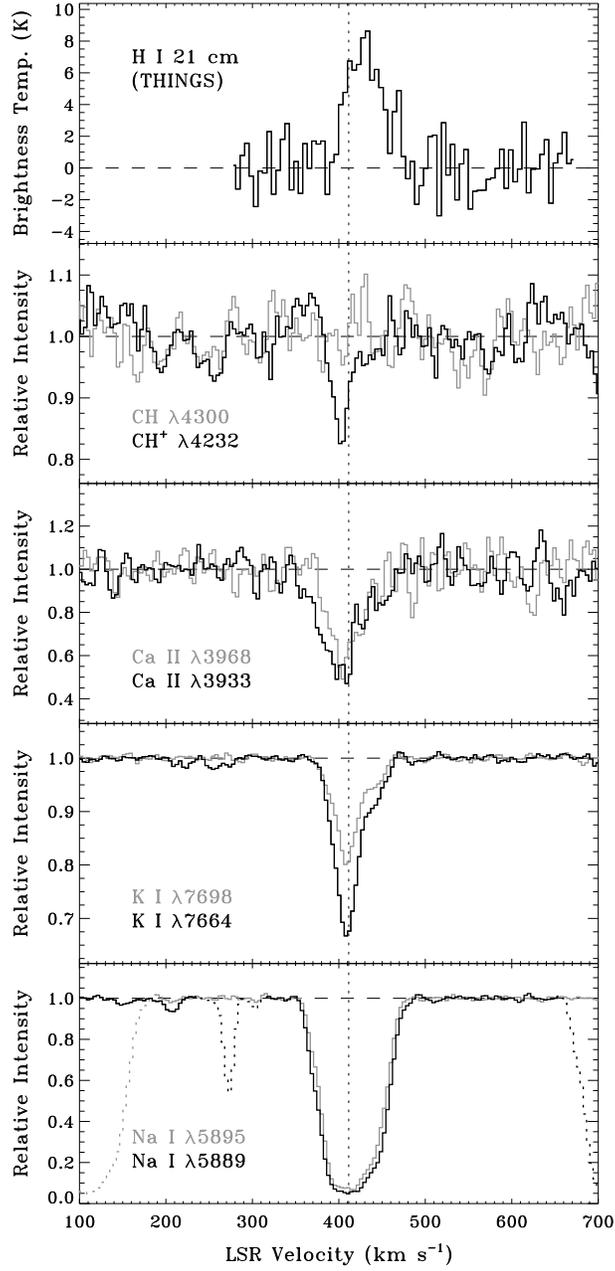}
\caption{Continuum-normalized interstellar absorption profiles of Na~{\sc i}, K~{\sc i}, Ca~{\sc ii}, and CH$^+$ in the direction of NGC~5195, along with the H~{\sc i}~21~cm emission profile of M51 at the position of NGC~5195 from THINGS. Spectra covering the CH region is also shown. Dotted features in the bottom panel denote overlap in the Na~{\sc i}~D lines. The vertical dotted line marks the mean column-density weighted velocity of the interstellar species seen in absorption ($v_{\mathrm{LSR}}=411$~km~s$^{-1}$). The mean H~{\sc i} velocity in this direction from the THINGS velocity field (moment 1 map) for M51 is $v_{\mathrm{LSR}}=434$~km~s$^{-1}$.}
\label{f3}
\end{figure}

\begin{figure}
\centering
\includegraphics[width=0.49\textwidth]{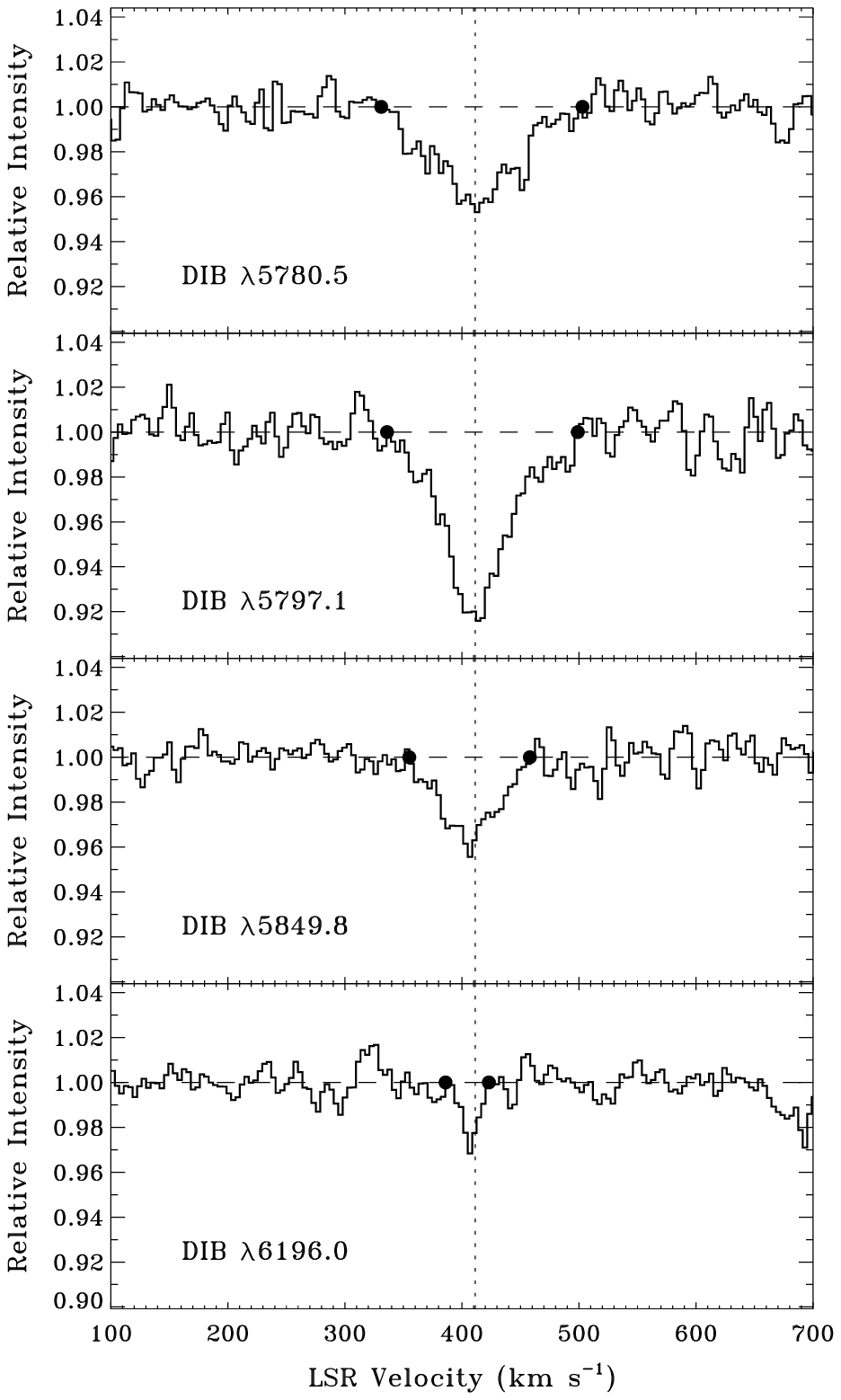}
\includegraphics[width=0.49\textwidth]{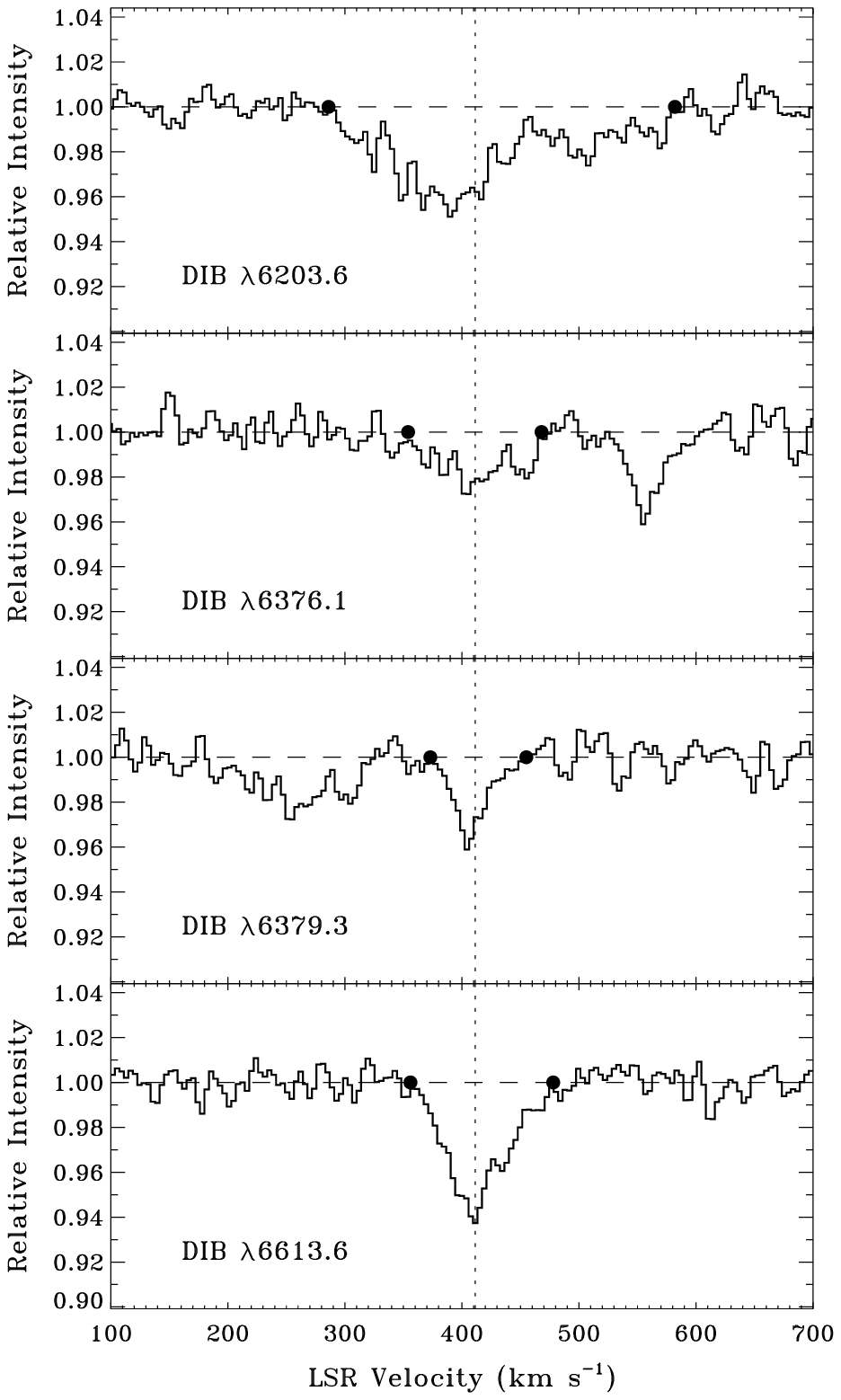}
\caption{Continuum-normalized absorption profiles of the eight DIBs detected in the direction of NGC~5195. The vertical dotted line marks the mean central velocity of the DIBs ($v_{\mathrm{LSR}}=411$~km~s$^{-1}$). Solid points indicate the limits of integration used to determine the equivalent widths. Note that the asymmetric profile of the $\lambda6203.6$ DIB, which includes an extended red wing, closely matches the profile shape of this DIB in Galactic sight lines (e.g., Friedman et al.~2011).}
\label{f4}
\end{figure}

\begin{figure}
\centering
\includegraphics[width=0.49\textwidth]{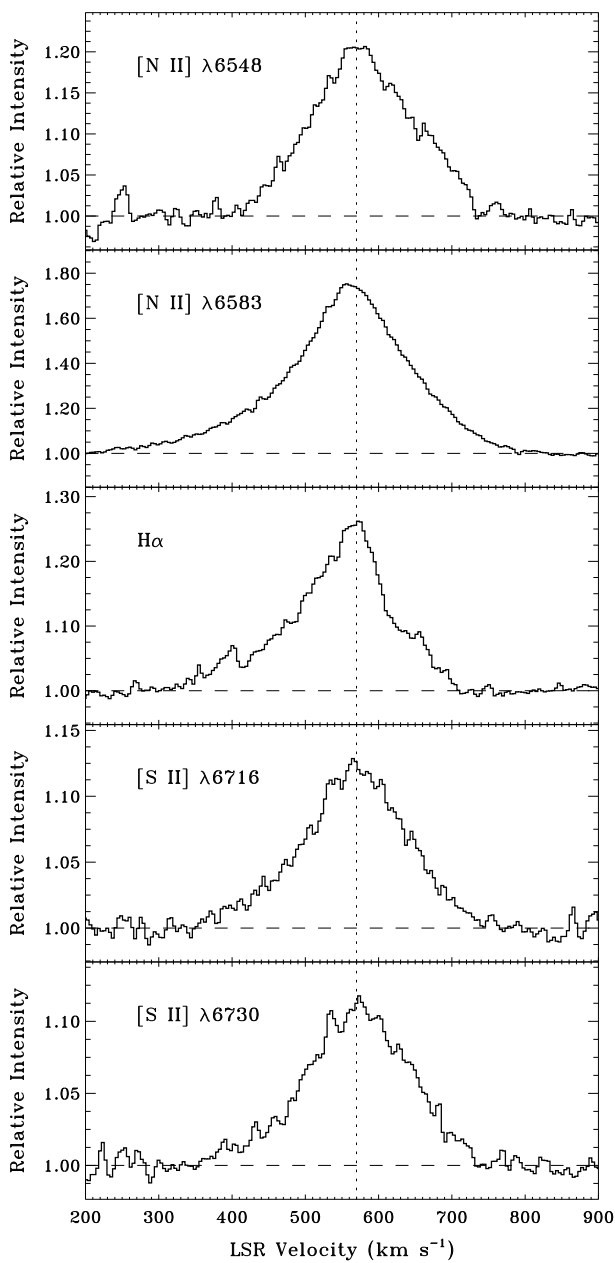}
\caption{Continuum-normalized emission profiles for prominent emission lines seen in the spectrum of NGC~5195. The continuum region surrounding the [N~{\sc ii}] and H$\alpha$ emission lines is affected by broad stellar H$\alpha$ absorption, which had to be fitted and removed from the spectrum before proceeding with the normalization. The vertical dotted line marks the mean centroid velocity of the emission lines listed in Table~1 ($v_{\mathrm{LSR}}=570$~km~s$^{-1}$).}
\label{f5}
\end{figure}

\begin{figure}
\centering
\includegraphics[width=0.49\textwidth]{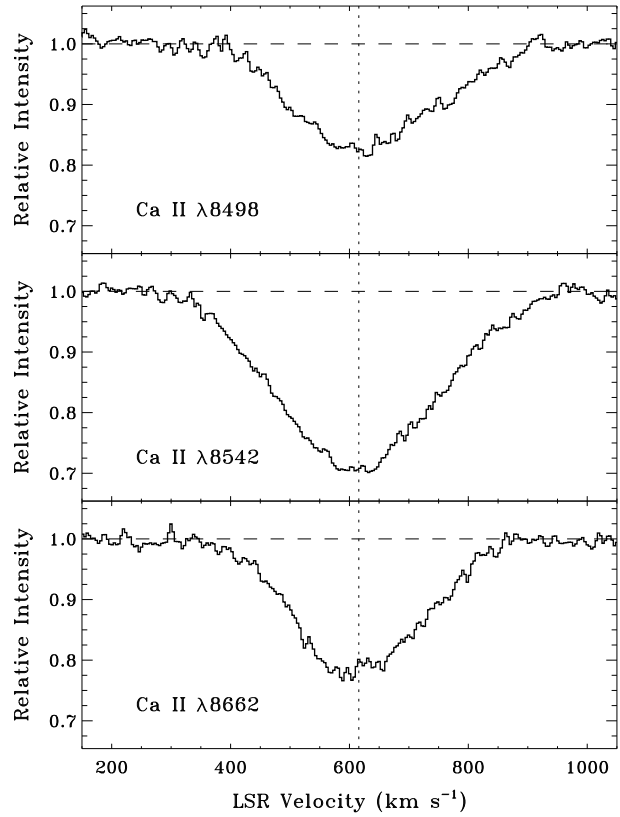}
\caption{Continuum-normalized absorption profiles for the Ca~{\sc ii} infrared triplet arising (presumably) from red giant stars in the nucleus of NGC~5195. The vertical dotted line marks the mean centroid velocity of the stellar absorption lines listed in Table~1 ($v_{\mathrm{LSR}}=616$~km~s$^{-1}$).}
\label{f6}
\end{figure}

\end{document}